\documentclass[10pt]{iopart}

\usepackage{graphicx}
\usepackage{iopams}
\expandafter\let\csname equation*\endcsname\relax
\expandafter\let\csname endequation*\endcsname\relax
\usepackage{amsmath}
\usepackage{cite}
\usepackage{bm}

\usepackage{lipsum,multicol}
\usepackage{multirow}

\begin{document}

\title{Boundary states in the chiral symmetric systems with a spatial symmetry}

\author{Jinpeng Xiao$^{1}$ and Jin An$^{1,2\dag}$}

\address{$^1$National Laboratory of Solid State Microstructures and Department of Physics, Nanjing University, Nanjing 210093, China}
\address{$^2$Collaborative Innovation Center of Advanced Microstructures, Nanjing University, Nanjing 210093, China}

\ead{$^{\dag}$anjin@nju.edu.cn}
\vspace{10pt}
\begin{indented}
\item[]October 2017
\end{indented}

\begin{abstract}
We study topological systems with both a chiral and a spatial symmetry which result in an additional spatial chiral symmetry. We distinguish the topologically nontrivial states according to the chiral symmetries protecting them and study several models in 1D and 3D systems. The perturbations breaking the spatial symmetry can break only one of the two chiral symmetries while the perturbations preserving the spatial symmetry always break or preserve both of them. In 3D systems, besides the 3D symmetries, the topologically nontrivial boundary modes may also be protected by the hidden lower dimensional symmetries. We then figure out the corresponding topological invariants and connect them with the 3D invariants.
\end{abstract}

%
\vspace{2pc}
\ioptwocol

\section{\label{level1}Introduction}

Topological materials have been blooming in recent years for they have many novel properties which may have promising applications in electronics industry. Symmetries play important roles in topological materials. The nonlocal symmetries such as time-reversal symmetry(TRS), particle-hole symmetry (PHS) and chiral symmetry influence the classifications of the topological systems and protect the nontrivial boundary states(NBSs) from perturbations\cite{Schnyder2008,Kitaev2009,Ryu2010}. Spatial symmetries have also attracted much attention recently, since they influence in a nontrivial way the classifications and properties of the topological materials.

For the chiral symmetric systems, chiral symmetry may come from sublattice structure in insulators\cite{Hosur2010,Essin2012,Wang2014},or from the coexistence of the intrinsic PHS and TRS\cite{Wakatsuki2014,Wakatsuki20142,Tewari2012,Tewari20122,Poyhonen2014,Xiao2015,Dumitrescu2015,Qu2015}, especially in superconductors. We study in this paper the chiral symmetric systems with an additional spatial symmetry. Although previous works have given the topological classifications and invariants of the systems in the presence of inversion symmetry\cite{Turner2010,Fu2007,Hughes2011,Lu2014,Engelhardt2015}, mirror symmetry\cite{Teo2008,Hsieh2012,Shiozaki2014,Chiu2013,Zhang2013,Ueno2013,Yoichi2015,Rusinov2016,Lau2016,Zhou2017,Hao2017}, or rotation symmetry\cite{Fu2011,Teo2013,Koshino2014}, they only focused on the subsystems such as the symmetric mirror planes or lines. Here we study an additional spatial symmetry for the whole system, which will give rise to another spatial chiral symmetry.

Each of the two chiral symmetries corresponds to a winding number, the nonzero value of which indicates the existence of the NBSs. In 1D $\mathbb{Z}$-characterized systems, we find a general method to distinguish the NBSs protected by a single chiral symmetry from that protected by both chiral symmetries. The NBSs protected by one chiral symmetry always contain the NBSs protected by the other chiral symmetry as shown in Figure \ref{fig_1}(a). The perturbations breaking the spatial symmetry always preserve one of the two chiral symmetries while the perturbations preserving the spatial symmetry always break or preserve both of them. In 3D systems, the hidden lower dimensional symmetries may also help to protect the nontrivial surface states.

This paper is organized as follows: In Sec.\ref{level3}, we study the $\mathbb{Z}$-characterized systems in 1D and 3D systems, respectively. Particular models are studied, and the processes of breaking one of the chiral symmetries and removing the corresponding NBSs are studied in detail. In Sec.\ref{level6}, we discuss the situations with multiple classifications. We summarize the results in Sec.\ref{level7}.

\section{\label{level3}$\mathbb{Z}$-valued topological systems}

In a topological system with Hamiltonian $H(k)$, the TRS $T$, PHS $C$ and chiral symmetry $S$ are defined as $TH(k)T^{-1}=H(-k)$, $CH(k)C^{-1}=-H(-k)$ and $\{H(k),S\}=0$. If there exists a spatial symmetry operator $\Gamma$ obeying $[H(k),\Gamma]=0$ and $\Gamma^{2}=1$, with $\Gamma$ commuting or anticommuting with $T$ or $C$, one can easily find a second set of symmetries with  $T'=\Gamma T$,  $C'=\Gamma C$ and $S'=\Gamma S$. Since the signs of squared operators of the TRS and PHS determine the topological classification\cite{Schnyder2008,Ryu2010}, and $T'^{2}(C'^{2})$ may not equal $T^{2}(C^{2})$, the classification of the system may be different by combining different TRS and PHS operators. For example, if $\Gamma$ anti-commutes with $T$($C$), the sign of $T'^{2}$($C'^{2}$) is opposite to $T^{2}$($C^{2}$), the classification of the system is thus changed by the spatial symmetry. The system then may belong to different topological classes simultaneously and may have multiple topological invariants. In the following, we first study the situation with definite classification in $\mathbb{Z}$-valued systems, and leave the brief discussion on that in $\mathbf{Z}_{2}$-valued systems in Appendix A. Then we discuss the situation with multiple classifications in the succeeding section.

\subsection{\label{level4} 1D systems}
In a $\mathbb{Z}$-valued topological 1D system of class such as AIII, BDI or CII, the topological invariant is defined as the winding number of the Hamiltonian $H(k)$\cite{Vayrynen2011}:
\begin{equation}\label{Eq2}
\begin{split}
\nu=\frac{1}{4\pi i}\int_{\mathrm{BZ}}dk\mathrm{Tr}[SH(k)^{-1}\partial_{k}H(k)],
\end{split}
\end{equation}
where $S$ is the chiral symmetry operator of the system.

\begin{figure}
\scalebox{1.0}{\includegraphics[width=0.5\textwidth]{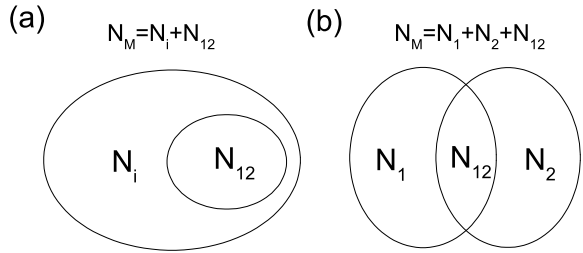}}
\caption{\label{fig_1} The schematic depictions of the possible relationships between different categories of NBSs in the system with two chiral symmetries $S$, $S'$. $N_{M}$ is the total number of the NBSs, $N_{1}$, $N_{2}$, $N_{12}$ are the numbers of NBSs protected by $S$, $S'$ and both, respectively. $N_{i}$ in (a) denotes $N_{1}$ or $N_{2}$. In real situation, only (a) can be realized while (b) never happens.}
\end{figure}
If the system has a spatial symmetry $\Gamma$ obeying $T\Gamma T^{-1}=\Gamma$ and $C\Gamma C^{-1}=\Gamma$, which indicates $[\Gamma,S]=0$ (for class AIII, only $[\Gamma,S]=0$ is needed), there exists a second chiral symmetry $S'=\Gamma S$. Correspondingly, there are two winding numbers $\nu$ and $\nu'$ related to $S$ and $S'$, respectively. The nontrivial values of the winding numbers mean the existence of the NBSs, and $\nu$($\nu'$) has the physical meaning of the number of the NBSs at each end protected by $S$($S'$). We demonstrate below in detail the relationship between the winding numbers and the numbers of the NBSs protected by a single chiral symmetry or both.

Since $\Gamma$ is a spatial symmetry of the whole system, the Hamiltonian can be block-diagonalized into subspaces labeled by the eigenvalues $\pm1$ of $\Gamma$. Assume the unitary transformation is $U$, then
\begin{equation}\label{Eq3}
\begin{array}{c c}
U^{\dag}\Gamma U=
\left(
\begin{array}{cc}
  1&   \\
   & -1  \\
\end{array}
\right),
U^{\dag}H(k)U=
\left(
\begin{array}{ccccc}
  H_{1}(k)&    \\
   & H_{2}(k)   \\
\end{array}
\right).
\end{array}
\end{equation}
Both $H_{1}(k)$ and $H_{2}(k)$ have the same classification as $H(k)$, and also preserve the chiral symmetry $S$, since $S$ is nonlocal.
The two winding numbers can then be written as $\nu=\nu_{1}+\nu_{2}$ and $\nu'=\nu_{1}-\nu_{2}$, where $\nu_{j}=\frac{1}{4\pi i}\int_{\mathrm{BZ}}dk\mathrm{Tr}[SH_{j}(k)^{-1}\partial_{k}H_{j}(k)]$ is the winding number for subsystem $j$, $j=1,2$.
Actually, all the NBSs can be divided into three categories: those only protected by one chiral symmetry $S$ or $S'$ and those protected by both. The numbers of them are denoted respectively as $N_{1}, N_{2}, N_{12}$. The total number of the NBSs of the system is denoted as $N_{M}$ and can be expressed as $N_{M}=\mid\nu_{1}\mid+\mid\nu_{2}\mid$. Thus we have
\begin{equation}\label{Eq4}
\begin{split}
&N_{M}=N_{1}+N_{2}+N_{12},\mid\nu\mid=N_{1}+N_{12},\\
&\mid\nu'\mid=N_{2}+N_{12}.
\end{split}
\end{equation}
Together with the expressions of $\nu$ and $\nu'$, we have
\begin{equation}\label{Eq05}
\begin{split}
&N_{M}=\frac{|\nu+\nu'|}{2}+\frac{|\nu-\nu'|}{2},N_{1}=N_{M}-|\nu'|,\\
&N_{2}=N_{M}-|\nu|, N_{12}=|\nu'|+|\nu|-N_{M}.
\end{split}
\end{equation}
Finally, we express $N_{1}, N_{2}, N_{12}$ in terms of $\nu$ and $\nu'$:
\begin{equation}\label{Eq5}
\begin{split}
&|\nu|>|\nu'|,\\
&N_{M}=|\nu|, N_{1}=|\nu|-|\nu'|,N_{2}=0, N_{12}=|\nu'|;\\
&|\nu|<|\nu'|,\\
&N_{M}=|\nu'|, N_{1}=0,N_{2}=|\nu'|-|\nu|, N_{12}=|\nu|.
\end{split}
\end{equation}
We show these results schematically in Figure \ref{fig_1}(a). What is remarkable is that the total number of the NBSs always equals the larger one of $|\nu|$ and $|\nu'|$, and there is no NBS protected only by the chiral symmetry relevant to the smaller one of $|\nu|$ and $|\nu'|$. This means there exists no system whose NBSs consist of three parts, where one part is protected only by $S$ ( $N_{1}\neq0$), one part only by $S'$ ($N_{2}\neq0$), while the remaining part is protected by both $S$ and $S'$ ($N_{12}\neq0$), as shown in Figure \ref{fig_1}(b).
We also find that $N_{1}=|\nu_{1}|+|\nu_{2}|-|\nu_{1}-\nu_{2}|=2m$, $N_{2}=|\nu_{1}|+|\nu_{2}|-|\nu_{1}+\nu_{2}|=2m'$, with $m,m'$ two integers, indicating that the number of NBSs protected only by one chiral symmetry is always even.
\begin{table}[!htp]
\caption{Possible perturbations and the chiral symmetry(symmetries) preserved by them. The first column indicates the commutation or anticommutation relationship between the perturbation and the spatial symmetry operator. The second to third columns give the perturbations with the subscripts indicating the commutation relationships with $T$ and $C$. For example, $\widetilde{\Omega}_{+-}$ means $T\widetilde{\Omega}T^{-1}=+\widetilde{\Omega}$, $C\widetilde{\Omega}C^{-1}=-\widetilde{\Omega}$. The fourth column gives the preserved chiral symmetry(symmetries).}\label{tab1}
\begin{tabular}{ccccc}
\hline
\hline
 & $\widetilde{\Gamma}_{\sigma,\sigma}$ & &$\widetilde{\Gamma}_{\sigma,\overline{\sigma}}$& CS\\
\hline
\multirow{2}{*}{$\{H_{p},\Gamma\}=0$}& $\widetilde{\Omega}_{\sigma,\overline{\sigma}},\widetilde{\Omega}_{\overline{\sigma},\sigma}$ &&$\widetilde{\Omega}_{\sigma,\sigma},\widetilde{\Omega}_{\overline{\sigma},\overline{\sigma}}$ & $S$ \\
&$\widetilde{\Omega}_{\sigma,\sigma},\widetilde{\Omega}_{\overline{\sigma},\overline{\sigma}}$& &$\widetilde{\Omega}_{\sigma,\overline{\sigma}},\widetilde{\Omega}_{\overline{\sigma},\sigma}$& $S'$\\
\hline
$[H_{p},\Gamma]=0$ &$\widetilde{\Omega}_{\sigma,\overline{\sigma}}$ & &$\widetilde{\Omega}_{\sigma,\sigma}$&$S,S'$ \\
\hline
\end{tabular}
\end{table}

Breaking chiral symmetry generically comes with removing the corresponding NBSs, which can be explicitly shown by the open-boundary band spectra. Now we study the chiral symmetry breaking processes by adding a perturbation without changing the classification of the system. Assume the perturbation takes the form $H_{p}\propto\widetilde{\Gamma}\widetilde{\Omega}$ with $\widetilde{\Gamma}$ the spatial and $\widetilde{\Omega}$ the nonspatial part. We classify all perturbations into two types, which consist of the spatial-symmetry breaking ones($\{\Gamma,H_{p}\}=0$) and preserving ones($[\Gamma,H_{p}]=0$), as shown in Table \ref{tab1}.
For spatial-symmetry breaking perturbations, they always preserve TRS and PHS, since $H_{p}$ can only break one of $T$ and $T'$(also $C$ and $C'$). So the system always has one couple of TRS and PHS, which means the system always has only one chiral symmetry.
For spatial-symmetry preserving perturbations, $[\Gamma,H_{p}]=0$, the TRS(PHS) operators $T(C)$ and $T'(C')$ have the same commutation relationships with $H_{p}$. The perturbation thus either holds the two chiral symmetries or breaks both. These results are shown explicitly in Table 1.

All the conclusions above are general for 1D chiral symmetric systems, including classes AIII, BDI and CII, except that in class CII, the $\mathbb{Z}$ invariants are actually $2\mathbb{Z}$ with $m$ and $m'$ even. We now confirm the above results by discussing below in detail one particular model in class BDI. The discussion can be extended straightforwardly to class AIII or CII.

We consider the recently studied Majorana bound states (MBSs) in helical arranged magnetic atomic chains on the surface of an s-wave superconductor\cite{Choy2011,Martin2012,Pientka2013,Nadj2013,Braunecker2013,Klinovaja2013,Vazifeh2013,Nakosai2013,Poyhonen2014,Kim2014,Pientka2014,Reis2014,Rontynen2014,Weststrom2015,Sedlmayr2015,Hu2015,Xiao2015}.
If all the magnetic atoms' momenta lie within a plane, the system has a chiral symmetry.
\begin{figure}
\scalebox{1.0}{\includegraphics[width=0.5\textwidth]{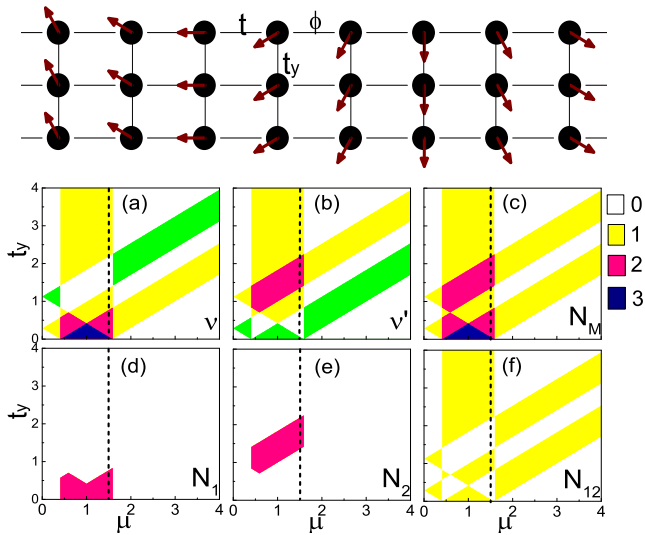}}
\caption{\label{fig2} The schematic depiction (upper panel) and the topological phase diagrams (lower panel) for the coupled triple chains of the magnetic atoms on an s-wave superconductor with helical magnetic structure. The pitch angle of each chain is $\phi$. The inter-chain coupling angle is 0. (a)((b))The topological invariant relevant to chiral symmetry $S$($S'$). (c)The total number of the MBSs. (d),(e) and (f) show the number of the MBSs protected only by $S$, $S'$ and both, respectively. Here $JS$=0.6, $\phi=2\pi/3$, $\Delta=0.1$, and the energy is in unit of $t$.}
\end{figure}
In polar coordinates, the momenta can be expressed as
$\mathbf{S}_{j}=S(\cos(j\phi),\sin(j\phi))$, with $j\phi$ the polar angle of $\mathbf{S}_{j}$.
After a spin-rotation transformation, the Hamiltonian for a $N$-chain ladder can be written in $k$-space as\cite{Xiao2015}
\begin{equation}\label{Eq6}
\begin{split}
H(k)=&H^{0}(k)-t_{y}\Gamma_{x}\tau_{z},\\
\end{split}
\end{equation}
where the coupling angle between the neighboring chains is 0 as shown in the upper panel of Figure \ref{fig2}. $H^{0}(k)$ is the single-chain Hamiltonian given by
\begin{equation}\label{Eq62}
\begin{split}
H^{0}(k)=&(-2t\cos\frac{\phi}{2}\cos k-\mu)\tau_{z}\\
&-2t\sin\frac{\phi}{2}\sin k\tau_{z}\sigma_{z}+JS\sigma_{x}+\Delta\tau_{x},\\
\end{split}
\end{equation}
where Pauli matrices $\tau, \sigma$ act on the particle-hole and spin spaces, and $t$ and $t_{y}$ are intra-chain and inter-chain hopping amplitudes. The matrices
\begin{equation}\label{Eq7}
\begin{array}{c c}
\Gamma_{x}=
\left(
\begin{array}{cccc}
  0&   1&    &      \\
  1&   0&   1&      \\
   &    .&   .&  . \\
   &    &    1&  0 \\
\end{array}
\right),
\Gamma_{y}=
\left(
\begin{array}{cccc}
  0&  -i&    &      \\
  i&   0&  -i&      \\
   &    .&   .&  . \\
   &    &    i&  0 \\
\end{array}
\right)
\end{array}
\end{equation}
are two $N\times N$ hermitian matrices acting on the inter-chain index space.
It is easy to find a spatial symmetry which is a reflection operator: $\Gamma=\left(\begin{smallmatrix}  & & & 1\\  & & 1& \\ ..&.& & \\  1 & & &  \end{smallmatrix}\right)$ and the chiral symmetries $S=\tau_{y}\sigma_{z}$ and $S'=\Gamma S=\Gamma\tau_{y}\sigma_{z}$.
The nonspatial TRS and PHS operators are $T=\sigma_{x}K$ and $C=\tau_{y}\sigma_{y}K$. The system belongs to class BDI, and the phase diagrams of the triple-chain case are shown in Figure \ref{fig2}.

\begin{figure}
\scalebox{1.0}{\includegraphics[width=0.5\textwidth]{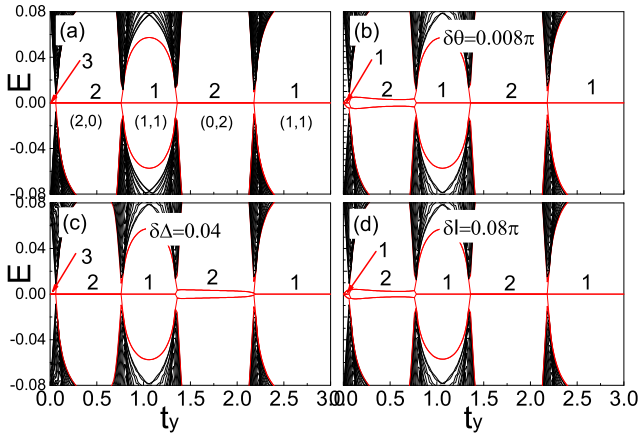}}
\caption{\label{fig3} The open-boundary spectra around zero energy (scan along the dash lines in Figure \ref{fig2}) for coupled triple chains with helical magnetic structure. (a) is for the unperturbed system with the coplanar structure and two chiral symmetries. The integer numbers below the flat bands in brackets denote ($\nu$,$\nu'$). (b)-(d) give the spectra for the system under perturbations (i),(ii),(iii) listed in the main text, respectively. The integer numbers above the flat bands denote the number of the MBSs or degeneracies of the non-zero-energy bands. Here $JS=0.6$, $\phi=2\pi/3$, $\Delta=0.1$ and $\mu=1.5$. The energy is in unit of hopping $t$, the magnitudes of perturbations are denoted in each subfigure.}
\end{figure}

Next, we study the chiral-symmetry-breaking effect of perturbations $H_{p}$ in this particular model.
For simplicity, we only study the on-site $\widetilde{\Gamma}_{++}$ perturbations in Table \ref{tab1}, i.e., we focus on the perturbations with their spatial parts commuting with both $T$ and $C$. All possible perturbations are shown in Table \ref{tab2}.
\begin{table}[!htp]
\caption{Possible perturbations and the chiral symmetry(symmetries) preserved by them. The first column indicates the commutation relationship between the spatial part of the perturbation and the spatial symmetry operator. The second column is the nonspatial part of the perturbation. The third column is the preserved chiral symmetry(symmetries).}\label{tab2}
\begin{tabular}{cccc}
\hline
\hline
\multirow{2}{*}{$\{\widetilde{\Gamma},\Gamma\}=0$}& $\sigma_{x},\sigma_{y},\tau_{x},\tau_{z},\tau_{x}\sigma_{z},\tau_{y}\sigma_{x},\tau_{y}\sigma_{y},\tau_{z}\sigma_{z} $ &  $S$ \\
&$\mathbb{I},\sigma_{z},\tau_{y},\tau_{x}\sigma_{x},\tau_{x}\sigma_{y},\tau_{y}\sigma_{z},\tau_{z}\sigma_{x},\tau_{z}\sigma_{y}$ &  $S'$\\

$[\widetilde{\Gamma},\Gamma]=0$ &$\sigma_{x},\sigma_{y},\tau_{x},\tau_{z}$ & $S,S'$ \\
\hline
\end{tabular}
\end{table}
For the triple-chain case, the spatial part of the perturbation is taken to be $\widetilde{\Gamma}=\Gamma_{z}^{1}=\left(\begin{smallmatrix}  1& & \\  & 0 & \\  &  & 1 \end{smallmatrix}\right)$ for perturbations preserving spatial symmetry and $\widetilde{\Gamma}=\Gamma_{z}^{2}=\left(\begin{smallmatrix}  1& & \\  & 0 & \\  &  & -1 \end{smallmatrix}\right)$ for those breaking spatial symmetry. For the former perturbation without changing the BDI classification, both chiral symmetries are preserved and any MBS cannot be removed. We now study the latter perturbation and consider three typical ones listed as follows:
\begin{enumerate}
\item $\delta \theta=\delta\theta_{1}=\pi-\delta \theta_{3}$, $JS\sigma_{x}\rightarrow JS\Gamma_{z}^{2}\sigma_{z}$;
\item $\delta\Delta=\delta\Delta_{1}=-\delta \Delta_{3}$, $\Delta\tau_{x}\rightarrow \Delta\Gamma_{z}^{2}\tau_{x}$;
\item $(\Delta_{1},\Delta_{2},\Delta_{3})=(\Delta e^{-i\delta I/2},\Delta,\Delta e^{i\delta I/2})$,$\Delta\tau_{x}\rightarrow\Delta\Gamma_{z}^{2}\tau_{y}$,
\end{enumerate}
where the subscript numbers are the chain indices and the dimensionless quantity $I$ in (iii) is
proportional to the supercurrent density along $y$ direction. Here perturbation (i) or (iii) breaks $S$, while perturbation (ii) breaks $S'$. The zero-energy MBSs may appear at the end of a topological superconductor. We show the open-boundary spectra of the unperturbed system with planar magnetic structure in Figure \ref{fig3}(a), where there are two chiral symmetries with $\mathbb{Z}$ invariants $\nu$ and $\nu'$. The total number of the MBSs is always directly related to the larger winding number. We divide the topological nontrivial parameter space into four regions denoted with ($\nu,\nu'$). The MBSs in region (2,0) and (0,2) are protected respectively only by $S$ and $S'$, while those in region (1,1) are protected by both. We add perturbations from (i) to (iii) listed above to observe the processes of removing part of the MBSs by breaking one of the two chiral symmetries. As expected, the perturbations only breaking $S$ remove the MBSs in region (2,0) in Figure \ref{fig3}(b)(d) while the perturbation breaking $S'$ removes the corresponding MBSs in region (0,2) in Figure \ref{fig3}(c).
However, the situations in 3D systems are quite different. We demonstrate them in next section.

\subsection{\label{level5} 3D system}
In 3D systems such as classes AIII, DIII and CI, the topological invariant is also characterized by a 3D winding number of the Hamiltonian $H(k)$ which can be calculated by\cite{Schnyder2008}
\begin{equation}\label{Eq11}
\begin{split}
\nu_{3D}=\frac{1}{48\pi^{2}}\int d^{3}\mathbf{k}\epsilon^{ijl}Tr[SH^{-1}\partial_{i}HH^{-1}\partial_{j}HH^{-1}\partial_{l}H].\\
\end{split}
\end{equation}
If there is a spatial symmetry $\Gamma$ in the system, and it commutes with symmetries $T$, $C$ and thus $S$ (in class AIII, only [$\Gamma,S$]=0 is needed), there exists a second chiral symmetry $S'=\Gamma S$. Therefore there exist two well defined winding numbers $\nu_{3D}$ and $\nu_{3D}'$ relevant to $S$ and $S'$. However, different to 1D systems where the winding numbers have the direct meaning of that of the NBSs, $\nu_{3D}$($\nu_{3D}'$) does not have the meaning of the number of Majorana/Dirac cones at the surface, so the NBSs cannot be classified in a similar way as in 1D systems. On the other hand, the NBSs at the surface of 3D systems are more robust than those in 1D systems. The reason is, for 3D systems, there also exists the lower dimensional hidden topological invariant\cite{Schnyder2011},
\begin{equation}\label{Eq13}
\begin{split}
\nu_{1D}=\frac{1}{4\pi i}\oint_{loop} dlTr[SH^{-1}\partial_{l}H],\\
\end{split}
\end{equation}
where the 1D loop is restricted in a small gapful area around a Majorana/Dirac cone located at the surface of the topological system. Actually, the Hamiltonian restricted to the loop belongs to class AIII, since the chiral symmetries are always preserved. Correspondingly, we also have two well defined 1D winding numbers for each surface cone: $\nu_{1D}$ and $\nu_{1D}'$ for $S$ and $S'$, either of which is +1 or -1. Thus, different from 1D systems, any single surface Majorana/Dirac cone of a 3D system is protected by both $S$ and $S'$. The 3D topological invariants $\nu_{3D}$ and $\nu_{3D}'$ are found to be the sums of all the winding numbers $\nu_{1D}$ and $\nu_{1D}'$ for the surface cones:
\begin{equation}\label{Eq14}
\begin{split}
\nu_{3D}=\sum\nu_{1D}, \nu_{3D}'=\sum\nu_{1D}'.\\
\end{split}
\end{equation}
Therefore the 3D winding numbers are not directly related to the numbers of cones.

\begin{figure}
\scalebox{1.0}{\includegraphics[width=0.5\textwidth]{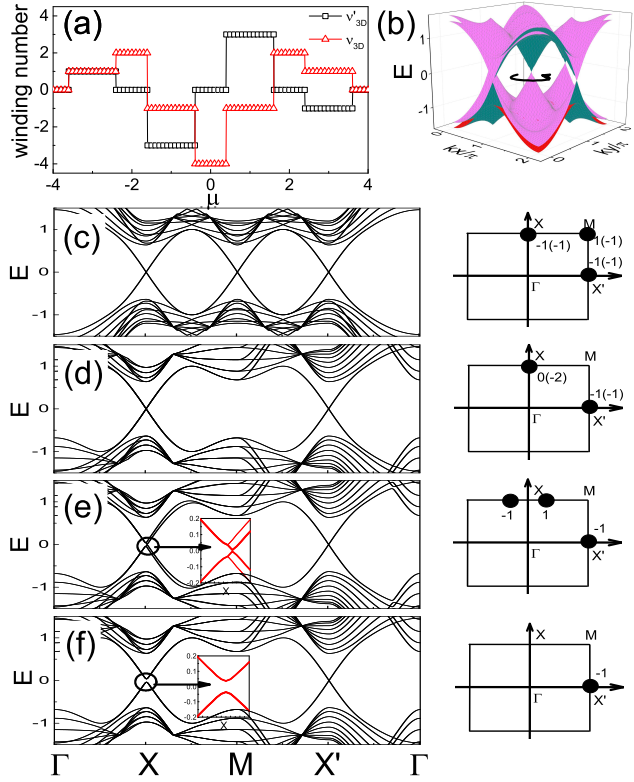}}
\caption{\label{fig4} The winding numbers and open-boundary spectra of a 3D topological superconductor with two chiral symmetries. (a)The two 3D winding numbers related to the two chiral symmetries when $m'=0.6$. (b)(c) show the surface Majorana cones located at three different high-symmetry positions on (001) surface. By moving the Majorana cone from M to X as shown in (d), a perturbation $0.05\Gamma_{x}\tau_{x}$ ($0.05\Gamma_{x}\tau_{x}\sigma_{x}$) which breaks $S'$ will split the degenerate cones into two non-degenerate single cones as shown in (e) (gap the cones as shown in (f)). Here $\mu=1$ in (b)-(f). The insets in (c)-(f) show the corresponding 2D surface Brillouin zones with Majorana points denoted with solid circles. The integer numbers near the points denote $\nu_{1D}(\nu_{1D}')$ in (c)(d) while denote $\nu_{1D}$ in (e)(f) since $S'$ is broken.   }
\end{figure}

As an example, we now start from a class DIII two-band topological superconductor, which can be obtained by doping a topological insulator\cite{Hosur2011}. The Hamiltonian can be written as
\begin{eqnarray}\label{Eq12}
H(k)&=
\left(
\begin{array}{cc}
 H_{0}(\mathbf{k})+m'\tau_{z}  & 0  \\
  0 &  H_{0}(\mathbf{k})-m'\tau_{z} \\
\end{array}
\right)\nonumber\\
&+a\Gamma_{j}\tau_{l}\sigma_{n},
\end{eqnarray}
where $H_{0}(\mathbf{k})=-(\mu+\sum_{i}\cos k_{i})\tau_{z}+\mathbf{d_{k}}\cdot\sigma\tau_{x}$ in Nambu basis $\Psi_{k}=[c_{k\uparrow},c_{k\downarrow},c_{-k\downarrow}^{\dag},-c_{-k\uparrow}^{\dag}]^{T}$, with $\mathbf{d_{k}}=(\sin k_{x},\sin k_{y},\sin k_{z})$, $i(l,n)=x,y,z$, while $j=x,y$. $\bm{\Gamma}$ are the Pauli matrices acting on the band space.
When $a=0$, the spatial symmetry is $\Gamma_z$, the two chiral symmetries are $S=\tau_{y}$ and $S'=\Gamma_z\tau_{y}$. Figure \ref{fig4}(a) shows the two 3D winding numbers relevant to the two chiral symmetries and Figure \ref{fig4}(b)-(c) explicitly show the surface Majorana cones when imposing open-boundary conditions along $z$ direction.

However, note that any non-degenerate cone at the high-symmetry points including $\Gamma(0,0)$, X$(0,\pi)$, M$(\pi,\pi)$ and X$'(\pi,0)$ is also protected by TRSs $(\Gamma_z)\sigma_{y}K$, since the 1D Hamiltonian $H(k_{x0},k_{y0},k_{z})$ as a function of $k_{z}$ belongs to class DIII, when $(k_{x0},k_{y0})$ is located at any of these points. Therefore each Majorana point of the non-degenerate cone acts as a pair of zero-mode Majorana fermions at the end of a 1D class DIII system, and this pair is thus protected by the TRSs since the 1D system is $\mathbf{Z}_{2}$ nontrivial. Since there always exists particle-hole symmetry in a superconductor, if a perturbation breaking both $S$ and $S'$ and thus also breaking the two TRSs is introduced, any non-degenerate cone can be gapped. However, if the introduced perturbation only breaks one of $S$ and $S'$, it thus also preserve one of TRSs, any non-degenerate cone cannot be gapped, since it is protected by the remaining chiral symmetry and TRS.

If two surface cones are located at the same position, they can be gapped without breaking both chiral symmetries. To make it clear, consider two surface cones at the same position with the opposite 1D winding numbers for $S$ but the same for $S'$. Although there still exist TRSs, the 1D Hamiltonian $H(k_{x0},k_{y0},k_{z})$ at this position becomes $\mathbf{Z}_{2}$ trivial since there are two pairs of zero modes at end. Without the protection from TRSs, the double cones are only protected by $S'$, since the total 1D winding number $\nu_{1D}'$($\nu_{1D}$) for $S'$($S$) of the loop around the double cones is $2$ or $-2$($0$), which is topologically nontrivial(trivial). Thus under a perturbation which breaks $S'$ but preserves $S$, the double cones can be gapped or split into two separated cones.

For example, we take $\mu=1$, and thus $\nu_{3D}=-1$, $\nu_{3D}'=3$, then at the surface $(001)$ there are three Majorana cones located respectively at X, M and X$'$, as shown in Figure \ref{fig4}(b)(c). The loop around a Majorana cone such as that shown in the inset arrow circle in Figure \ref{fig4}(b) belongs to class AIII, and thus has the 1D winding number +1 or -1 regardless with respect to $S$ or $S'$. The three Majorana cones are found to be robust against all kinds of perturbations without breaking both chiral symmetries. However, if we move the cones located at M($\pi,\pi$) and X($0,\pi$) together, the perturbations coupling the two cones and simultaneously breaking chiral symmetry $S'$ may gap or split them. To create double surface cones without breaking any symmetry, we modify the Hamiltonian as follows
\begin{eqnarray}\label{Eq14}
H(k)&=
\left(
\begin{array}{cc}
 H_{0}'(\mathbf{k})+m'\tau_{z}  & 0  \\
  0 &  H_{0}(\mathbf{k})-m'\tau_{z} \\
\end{array}
\right)\nonumber\\
&+a\Gamma_{j}\tau_{l}\sigma_{n},
\end{eqnarray}
where $H_{0}'(\mathbf{k})=H_{0}(k_{x}-\pi,k_{y},k_{z})$.
In Figure \ref{fig4}(d)-(f), we show the processes of gapping or splitting Majorana cones. The loop integral around point X is -2 with $S'$ but 0 with $S$, which means the two Majorana cones at X are only protected by $S'$. As can be seen in Figure \ref{fig4}(e), the perturbation breaking $S'$ splits them into two separated cones away from X. Each split cone is protected by $S$ and the winding numbers are +1 and -1, respectively. While in Figure \ref{fig4}(f), another perturbation breaking $S'$ gaps the cones. The results can be extended to a 2$\mathbb{Z}$-characterized class CI system, while the only difference is that its 3D winding numbers and the number of Majorana surface cones are always even.

\section{\label{level6} The situations with multiple classifications}
We have discussed the situation where the spatial symmetry operators commute with the TRS and PHS. There are still some situations where the spatial symmetry operators anti-commute with the TRS, PHS or both.

In class DIII, BDI, CI or CII, if only one of TRS and PHS commutes with $\Gamma$, the chiral symmetry has $\{\Gamma,S\}=0$(for class AIII, only $\{\Gamma,S\}=0$ is needed), the winding numbers will always be zero. The explanation is as follows: under the transformation which block-diagonalizes $\Gamma$, $H(k)$ is also block-diagonalized. However, $S$ is off-diagonalized as $\left(\begin{smallmatrix}  0 & D\\  D^{\dag}& 0\end{smallmatrix}\right)$, then the traces in Eq. (\ref{Eq2}) and Eq. (\ref{Eq11}) are zero, so are the integrals relevant to both chiral symmetries. In such situations, the system has multiple classifications. For example, if $T\Gamma T^{-1}=-\Gamma$ and $C\Gamma C^{-1}=\Gamma$ in an original class BDI system, the system can be in class BDI and DIII simultaneously as $T^{2}=+1$ and $T'^{2}=-1$. The system may have MBSs but are irrelevant to the winding numbers.

If $[\Gamma,S]=0$ and $\Gamma$ anti-commutes with both of the TRS and PHS, this additional symmetry will make the system have quadruple classifications.
Consider a class DIII topological superconductor, there are TRS $T$($T^{2}=-1$), PHS $C$($C^{2}=+1$), chiral symmetry $S$ and spatial symmetry $\Gamma$. We also have the second TRS $T'=\Gamma T$, PHS $C'=\Gamma C$ and spatial chiral symmetry $S'=\Gamma S$. If $\{T,\Gamma\}=0$ and $\{C,\Gamma\}=0$, then $T'^{2}=+1, C'^{2}=-1$.
The classifications will be
\begin{enumerate}
\item $T, C, S$,  class DIII;
\item $T, C', S'$, class CII;
\item $T', C, S'$, class BDI;
\item $T', C', S$,  class CI.
\end{enumerate}

Generally, the topological classification is made for minimal systems without additional symmetry. Therefore, the model Hamiltonian with a spatial symmetry is not minimal at all and it can always be block-diagonalized. However, the multiple classifications and multiple topological invariants are still valid. Actually, one can consider the gapful system in a block-diagonalized representation. Without closing the bulk gap, one can always add a small off-diagonal perturbation to break some symmetries to make three out of the four classifications invalid, leaving only one classification survived. Thus with the perturbation the model Hamiltonian becomes minimal and its corresponding topological invariant is well defined. When adiabatically removing the perturbation, the invariant always keeps well defined, since the bulk gap is open. Similarly, the other three classifications and their corresponding topological invariants are well defined. It is then meaningful that when the system has an additional symmetry, it simultaneously belongs to multiple classes and has multiple topological invariants which are related to each other.

In 1D systems, they are characterized by $\mathbf{Z}_{2}$, 2$\mathbb{Z}$, $\mathbb{Z}$ and none, respectively. Both the classifications (ii) and (iii) are characterized by the winding number $\nu'$ which is relevant to $S'$ and must be even, while the $\mathbb{Z}_{2}$ invariant $W=(-1)^{\nu'/2}$ is actually the parity of the Kramers pairs. The NBSs protected by the $\mathbf{Z}_{2}$ invariant are always protected by the $\mathbb{Z}$ characterized $S'$. For example, in Hamiltonian (\ref{Eq6}), if the inter-chain coupling angle is $\pi$, and the chain number is even, the last term will become $-t_{y}\Gamma_{y}\tau_{z}\sigma_{z}$. The spatial symmetry operator will be an antisymmetric matrix $\Gamma'=\left(\begin{smallmatrix}   &&&-i\\  &&i&\\ &...&&\\i &&&\end{smallmatrix}\right)$, which anticommutes with both $T$ and $C$, resulting in quadruple classifications for the multi-chain system.

In 2D systems, there exist only one class DIII which has nontrivial topological $\mathbf{Z}_{2}$ invariant, where all the NBSs are related to $S$.

In 3D systems, they are characterized by $\mathbb{Z}$, $\mathbf{Z}_{2}$, none and 2$\mathbb{Z}$, respectively. Here both the classifications (i) and (iv) are characterized by the winding number $\nu$ which is relevant to $S$ and must be even. This indicates that the number of Majorana cones is also even, since the 1D winding number for each cone is $\pm1$. On the other hand, the $\mathbf{Z}_{2}$ invariant for class CII characterize the parity of the number of pairs of these Majorana cones.

\section{\label{level7}Summary}

We have studied the chiral symmetric systems with two chiral symmetries. The additional spatial chiral symmetry comes from combining the spatial symmetry and the nonspatial chiral symmetry. When the spatial symmetry operator commutes with the nonlocal antiunitary symmetry operators, the topological classification is not changed.
We found methods to distinguish the nontrivial boundary states according to the chiral symmetries protecting them.

There are two winding numbers in the $\mathbb{Z}$-valued systems in 1D and 3D which are relevant to the two chiral symmetries.
In 1D systems, we classified the nontrivial boundary states into three categories where two of them are protected respectively by a single chiral symmetry and the remaining one is protected simultaneously by both chiral symmetries. We found the relationship between the winding numbers and the numbers of nontrivial boundary states protected by each chiral symmetry and confirmed the results by adding perturbation to break chiral symmetries. When breaking one of the two chiral symmetries, only the corresponding boundary states are removed.

In 3D systems, we found the 3D winding numbers do not directly indicate the numbers of nontrivial surface cones protected by the chiral symmetries. There are two lower dimensional 1D winding numbers around the gapless surface cones. For each chiral symmetry, the 3D winding number is the sum of the corresponding lower dimensional 1D winding numbers. The 1D winding numbers for each single cone are always ``+1'' or ``-1'' for the two chiral symmetries. Thus any non-degenerate cone is protected by both chiral symmetries and robust against any perturbation which breaks only one chiral symmetry. However, if the surface cones are degenerate, breaking one chiral symmetry may split or gap them.

We also have studied the situations with multiple classifications. When the spatial symmetry operator anti-commutes with both nonlocal antiunitary symmetry operators, the systems' classifications are quadruple and characterized by $\mathbb{Z}$ and $\mathbf{Z}_{2}$ invariants simultaneously.

\section*{Acknowledgements}
This work was supported by NSFC under grant No.11174126, and the State Key Program for Basic Researches of China under grant No.2015CB921202.

\section*{\label{level8}Appendix A. $\mathbf{Z}_{2}$-valued topological systems}

Kitaev has proposed a topological $\mathbf{Z}_{2}$ invariant relevant to PHS in 1D spinless p-wave superconductor system\cite{Kitaev2001} with TRS($T^{2}=+1$). However, the systems with TRS($T^{2}=-1$) always can be valued by a $\mathbf{Z}_{2}$ invariant which is relevant to the parity of Kramers pairs. This basic topological invariant even exists in the systems with additional symmetries, although a more comprehensive $\mathbb{Z}$ valued invariant may appear.

We introduce the Kane-Mele $\mathbf{Z}_{2}$ invariant\cite{Kane2005}, where $W=\prod_{K_{0}}\frac{Pf[w(K_{0})]}{det[w(K_{0})]}$, $K_{0}$ denotes TRS points in $k$ space and $w_{nm}(k)=<u_{n}(-k)\mid T\mid u_{m}(k)>$, with $\mid u_{n}(k)>$ the Bloch state for $n$th occupied band of Hamiltonian with TRS. If there is a spatial symmetry $\Gamma$ which does not change the topological classification of the system, we have $T^{2}=T'^{2}=-1$. For simplicity, we block diagonalize the Hamiltonian $H(k)=diag\{H_{1}(k),H_{2}(k)\}$ and $\Gamma=\Gamma_{z}=diag\{1,-1\}$, so $T'=\Gamma_{z}T$. Then it's easy to find that $w(k)=diag\{w_{1}(k),w_{2}(k)\}$, $w(k)'=diag\{w_{1}(k),-w_{2}(k)\}$. Because the TRS points must appear in pairs, $\prod_{K_{0}}Pf[w(K_{0})]=\prod_{K_{0}}Pf[w(K_{0})']$ and $\prod_{K_{0}}det[w(K_{0})]=\prod_{K_{0}}det[w(K_{0})']$, $W=W'$. The $\mathbf{Z}_{2}$ invariants calculated by two TRS operators are exactly the same, which means any topologically nontrivial boundary state is equally protected by both TRSs.

\section*{References}
\bibliography{refs}

\end{document}